\documentclass{sig-alternate-05-2015}

\usepackage{tabularx}
\usepackage{tabulary}
\usepackage{booktabs}
\usepackage{datetime}
\usepackage[show]{notes-alt}
\usepackage{subfigure}
\usepackage[ruled,vlined,linesnumberedhidden]{algorithm2e}

\usepackage{float}
\usepackage{url}
\usepackage{tabto}
\usepackage{amsmath}
\usepackage{enumitem}

\usepackage[numbers,square]{natbib}

\begin{document}

\CopyrightYear{2016} 
\setcopyright{acmcopyright}
\conferenceinfo{JCDL '16,}{June 19-23, 2016, Newark, NJ, USA}
\isbn{978-1-4503-4229-2/16/06}\acmPrice{\$15.00}
\doi{http://dx.doi.org/10.1145/2910896.2910901}

\newcommand{\seq}[1]{\langle\,{#1}\,\rangle}
\newcommand{\set}[1]{\left\{\,{#1}\,\right\}}
\newcommand{\bigset}[1]{\left\{\,{#1}\,\right\}}
\newcommand{\bigtuple}[1]{\left(\,{#1}\,\right)}
\newcommand{\card}[1]{\left|\,{#1}\,\right|}
\newcommand{\tup}[1]{\left(\,{#1}\,\right)}
\newcommand{\argmin}[2]{\underset{#1}{\operatorname{arg\,min}}{\:\: #2}}
\newcommand{\argmax}[2]{\underset{#1}{\operatorname{arg\,max}}{\:\: #2}}
\newcommand{\stdmin}[2]{\underset{#1}{\operatorname{min}}{\:\: #2}}
\newcommand{\simplemin}[1]{\ensuremath{\underset{#1}{min}\;}} 
\newcommand{\pow}[2]{\ensuremath{#1^#2\;}}
\newcommand{\ra}[1]{\renewcommand{\arraystretch}{#1}}

\hyphenation{Map-Reduce}
\hyphenation{opti-mi-za-tion}

\title{The Dawn of Today's Popular Domains}
\subtitle{A Study of the Archived German Web over 18 Years\thanks{This work is partly funded by the European Research Council under ALEXANDRIA (ERC 339233)}}

\numberofauthors{1}
\author{
\alignauthor
Helge Holzmann, Wolfgang Nejdl, Avishek Anand\\
       \affaddr{L3S Research Center}\\
       \affaddr{Appelstr. 9a}\\
       \affaddr{30167 Hanover, Germany}\\
       \email{\{holzmann,nejdl,anand\}@L3S.de}
}

\maketitle

\begin{abstract}
The Web has been around and maturing for 25 years. The popular
websites of today have undergone vast changes during this period, with
a few being there almost since the beginning and many new ones
becoming popular over the years. This makes it worthwhile to take a
look at how these sites have evolved and what they might tell us about
the future of the Web. We therefore embarked on a longitudinal study
spanning almost the whole period of the Web, based on data collected
by the Internet Archive starting in 1996, to retrospectively analyze
how the popular Web as of now has evolved over the past 18 years.

For our study we focused on the German Web, specifically on the top
100 most popular websites in 17 categories. This paper presents a
selection of the most interesting findings in terms of \emph{volume},
\emph{size} as well as \emph{age} of the Web. While related work in
the field of Web Dynamics has mainly focused on change rates and
analyzed datasets spanning less than a year, we looked at the
evolution of websites over 18 years. We found that around 70\% of the
pages we investigated are younger than a year, with an observed
exponential growth in age as well as in size up to now. If this growth
rate continues, the number of pages from the popular domains will
almost double in the next two years.  In addition, we give insights
into our data set, provided by the Internet Archive, which hosts the
largest and most complete Web archive as of today.
\end{abstract}

%
%

\ccsdesc[500]{Information systems~World Wide Web}
\ccsdesc[500]{Applied computing~Digital libraries and archives}

\printccsdesc

\keywords{Web Dynamics; Analysis; Statistics; Longitudinal; Retrospective}

\section{Introduction}
\label{sec:introduction}

The Web is in a state of continuous change, with websites and pages being continuously added, deleted and modified. As previous studies have reported, the Web has been growing and evolving substantially over its lifetime. Researchers have measured and characterized the nature and degree of change in the past~\cite{cho_evolution_1999,fetterly_large-scale_2003,koehler_web_2002,ntoulas_whats_2004,adar_web_2009}. However, these studies primarily focus on content or structural change rates of rather small collections of websites for time periods from a few weeks to a couple of years. One of the interesting findings of such analyses is that a significant part of the changes on the Web are the creation and deletion of pages \citep{ntoulas_whats_2004}. This paper aims to extend those studies with a comprehensive retrospective analysis with a strong topicality, by investigating today's most prominent part of the German Web over an 18 year period from 1996 to 2013. We collected the most popular domains from a diverse set of categories on Amazon's \textit{Alexa} ranking\footnote{\url{http://www.alexa.com}} and analyzed on the German Web crawls for this period preserved by the \textit{Internet Archive}\footnote{\url{http://www.archive.org}}. This makes it the longest study of Web evolution so far.

The dataset gives us the unique opportunity to analyze the evolution of what is popular on the Web today and how those websites have evolved from their early days. At the same time it puts us into a role similar to an archaeologist, who studies the past only based on what has remained. What remains of the Web in archives is influenced by crawling policies, which are limited due to the available resources. Furthermore, not only the Web itself but also the crawlers are subject to evolution. Therefore, we will discuss our assumptions and findings on the Internet Archive dataset in a separate section, which by itself is another interesting contribution of the study described in this paper and shows the representativeness of the archive with respect to the most popular websites by comparing the growth to the actual Web in terms of registered domains. In this respect, it is interesting to see that those websites are relatively well covered, even though some years back they might not have been as popular as today. This is a positive observation and an important trait of a Web archive since today's popular websites are likely to be looked up by users of an archive from the past as well.

\begin{sloppypar}
In the following we will use \textit{domain} synonymously for a \textit{website} including its \textit{sub-domains}, e.g., \textit{google.de} and \textit{news.google.de} belong to one website. In contrast, \textit{webpage} is used interchangeably for \textit{URL} and denotes a single \textit{page} of a website.
\end{sloppypar}

The questions we ask in our study are inspired by the popular belief about the structure of the Web, but with a focus on the prominent part that people care about most today in Germany:
\begin{itemize}[noitemsep]

	\item \emph{Are popular websites growing old and if so, how can we characterize it?} We were able to confirm what other researchers found earlier: the majority of pages on the Web are rather young. In addition, however, we found that the small long-living fraction contributes significantly to the age, which is increasing.

	\item \begin{sloppypar}\emph{How has the size of popular websites changed over time? } In terms of the volume of a domain, which we define as the number of URLs, we found the growth has been exponential up to now. This is an interesting finding, which we believe is true for the Web in general. Regarding actual sizes, not just existing pages grow, but also newly created ones are larger every year.\end{sloppypar}

	\item \emph{Do the popular websites from different categories (like business, universities and technology) have different growth rates?} In almost all the conducted analyses we found distinct differences among the considered categories. We find that 75\% of the popular university domains of today have been around since \emph{1999} whereas not even 20\% of the popular game websites of today were present back then.

\end{itemize}

Before we present the results of our analysis (Sec.~\ref{sec:age-analysis} and~\ref{sec:url-analysis}), we provide a detailed description of the experimental setup and the measurement metrics used in this study (Sec.~\ref{sec:setup}). Since all presented properties and statistics are computed purely on meta data from a crawl index (CDX), the same analysis can be replicated by other researchers with access to such an index. Using the same definitions (Table~\ref{tab:property_reference}) would allow to compare among datasets, e.g., different national domains. The national top-level domain \textit{.de} constitutes the largest fraction of German-speaking websites, a non-negligible portion of the Web, which we analyze with a focus on the most popular part. The paper ends with an analysis and discussion of this dataset as provided by the \textit{Internet Archive} (Sec.~\ref{sec:dataset_analysis}).

\section{Related Work}
\label{sec:related-work}

Studying and characterizing change and evolution in the Web falls into the broad field of \emph{Web Dynamics}. Change on the Web can be differentiated into content change and structural change in terms of the Web graph as well as the creation and deletion of webpages. This paper investigates the latter together with the growth of Web pages as a result of content change, which is not analyzed in depth though, as we operated purely on metadata.

By contrast, the earliest studies in this field mainly investigated content changes with respect to change rates. Already in 2000, \citet{cho_evolution_1999} analyzed 720,000 pages over 4 months in a study motivated by the question on \textit{how to build an effective incremental crawler}. They found that 40\% of them change within a week based on their checksum. Similar to us, they focused on popular pages, determined by computing PageRank. In a similar study from 2003, \citet{fetterly_large-scale_2003} analyzed 150 million webpages over a period of 11 weeks with more sophisticated features. They found that 67\% of the pages never change, 20\% are only minor text changes and 10\% of the webpages have changes in the non-textual part. Only around 4\% of the webpages report medium to major changes to their text content. The first study in this respect that covers multiple years was done by \citet{koehler_web_2002} in 2002. They analyzed a small sample of 360 pages spanning more than four years from 1996 to 2001 and showed that navigation pages have a better survival rate than content pages. A more fine-grained content analysis was done much later by \citet{adar_web_2009} in 2009, taking hourly and sub-hourly changes into account. They studied page level content changes and tried to capture term-level dynamics on a sample of 55,000 pages with different popularities and different revisitation patterns over 5 weeks. They found that 66\% of the visited pages changed during the period under consideration on average every 123 hours.

From a search engine perspective, back in 2004, \citet{ntoulas_whats_2004} analyzed the link structure in addition to content of 3-5 million pages over one year. They focused on popular websites once again, according to Google's directory, and observe that 8\% of the pages are replaced by newly created ones every week. Out of the remaining about 50\% did not change at all during the year under consideration.

With a focus purely on structural change, \citet{baeza2006} investigated the Chilean Web (\textit{.cl}) domain over five years from 2000 to 2004 with questions similar to ours. During this period, their collection grew from 600,000 to 3 million pages. Other studies also focused on national top-level domains, such as \textit{.uk}, which was studied by \citet{bordino2008} in 2008 as well as in a recent study from 2014 by \citet{hale_mapping_2014}. \citet{bordino2008} analyzed a time-aware Web graph consisting of 100 million pages over one year with monthly granularity. \citet{hale_mapping_2014} focused on the academic part of the UK under \textit{.ac.uk} from 1996 to 2010 and investigated link patterns. As in our study their collection was also crawled and provided by the Internet Archive. Another recent work by \citet{agata:2014} analyzed a collection of 10 million mainly Japanese pages in 2001, which was collected by the Internet Archive as well and is also based on metadata. They report a webpage's average life span of a little more than three years. The most recent study with a national focus was published by \citet{alkwai2015} in 2015. They analyzed around 300,000 Arabic pages in terms of different criteria, such as their coverage on Web archives.

Our work differs from these previous analyses by having a larger temporal coverage as well as new objectives. To this effect, we carry out studies which compare observations across years showcasing evolution of websites in terms of age (s. Sec~\ref{sec:age-analysis}) and growth both in size and volume (s. Sec~\ref{sec:url-analysis}).

\section{Setup and Methodology}
\label{sec:setup}

\begin{table}[!t]
  \small
  \centering  
  \caption{Dataset Details}
\begin{tabular}{rrrr}
    \toprule
    \textbf{Category}&\textbf{\#~Domains}&\textbf{\#~Sub-Domains}&\textbf{\#~URLs}\\
    \midrule
    Computer & 100 & 561 & 2138786\\
    Recreation & 100 & 380 & 981638\\
    Society & 100 & 368 & 832017\\
    Health & 100 & 274 & 453282\\
    Kids~\&~Teens & 100 & 234 & 311705\\
    Culture & 100 & 250 & 934552\\
    Media & 100 & 512 & 1981877\\
    Shopping & 100 & 429 & 6726195\\
    Regional & 100 & 793 & 3069791\\
    Games & 99 & 304 & 718348\\
    Sports & 100 & 290 & 656859\\
    Business & 100 & 546 & 1534639\\
    Education & 100 & 827 & 1240196\\
    Science & 100 & 398 & 579821\\
    Home & 100 & 325 & 1762361\\
    News &  40 & 117 & 820163\\
    Universities & 100 & 828 & 659175\\
    \hline
    \textbf{\textit{TOTAL}} & \textbf{\textit{1444}} & \textbf{\textit{5846}} & \textbf{\textit{20778475}}\\
    \bottomrule
    \vspace{-0.5cm}
  \end{tabular}
  \label{tab:dataset_statistics}
\end{table}

\begin{table*}[!t]
  \small
  \centering  
  \caption{Properties Used in the Statistics}
  \begin{tabular}{ll}
    \toprule
    \multicolumn{2}{l}{\textit{Evolution} and \textit{Domain Age} statistics}\\
    \midrule
    
\textbf{$\text{alive}_d(p_i)$} & \# URLs of $d$ alive in period $p_i$ (were born before $t_i$ and did not die before $t_{i+1}$)\\

\textbf{$\text{born}_d(p_i)$} & \# URLs of $d$ born in period $p_i$
(were born after $t_i$ (included) and did not die before $t_{i+1}$)\\

\textbf{$\text{died}_d(p_i)$} & \# URLs of $d$ died in period $p_i$
(were born before $t_i$ and died before $t_{i+1}$)\\

\textbf{$\text{flashed}_d(p_i)$} & \# URLs of $d$ born and died in period $p_i$
(were born after $t_i$ (included) and died before $t_{i+1}$)\\

\textbf{$\text{size}_d(p_i)$} & Cumulated sizes of URLs of $d$ at the end of period $p_i$
(all URLs that were alive or were born in period $p_i$)\\

\textbf{$\text{born\_size}_d(p_i)$} & Cumulated sizes of URLs of $d$ at the birth of newborn URLs in period $p_i$\\

\textbf{$\text{ages}_d(p_i)$} & Ages in months of URLs of $d$ at the end of period $p_i$
(all URLs that were alive or were born in period $p_i$)\\

\midrule
\multicolumn{2}{l}{\textit{URL Age} statistics}\\
\midrule

\textbf{$\text{count}_d(p_i)$} & \# URLs of $d$ in period $p_i$ / at age $i$
(were born before $t_i$ and reached age $i$)\\

\textbf{$\text{died}_d(p_i)$} & \# URLs of $d$ that died in period $p_i$ / at age $i$
(were born before $t_i$ and died before $t_{i+1}$)\\

\textbf{$\text{size}_d(p_i)$} & Cumulated sizes of URLs of $d$ at the end of period $p_i$
(only of URLs that did not die in period $p_i$)\\

\textbf{$\text{died\_size}_d(p_i)$} & Cumulated sizes at the death of URLs of $d$ that died in period $p_i$\\

\textbf{$\text{died\_birth\_size}_d(p_i)$} & Cumulated sizes at the birth of URLs of $d$ that died in $p_i$\\

    \bottomrule
    \vspace{-0.5cm}
  \end{tabular}
  \label{tab:property_reference}
\end{table*}

In our analysis we focused on the aging as well as growth of today's most popular German websites based on a Web archive over 18 years. All information needed for such an analysis are available in the meta data index, called CDX\footnote{\url{http://archive.org/web/researcher/cdx_file_format.php}}, which most Web archives maintain with their collections.

\subsection{Dataset Preparations}

Our dataset has been provided by the Internet Archive in the context of the ALEXANDRIA project\footnote{\url{http://alexandria-project.eu}} and consists of
all their archived text records from the German Web, as defined by the
\textit{.de} top-level domain, from 1996 to 2013.

\subsubsection{German Web CDX}
\label{sec:cdx}

The so-called CDX files that we used for our investigation are manifests consisting of all meta information about the crawls in a space-separated format, with one line per capture, i.e. a snapshot of one URL at a given time. The corresponding line in the CDX file looks as follows:

\texttt{ <canonicalized\_url\hspace{12pt}timestamp\hspace{12pt}original\_url\\mime\_type\hspace{12pt}status\_code\hspace{12pt}checksum\hspace{12pt}redirect\_url\\meta\_data\hspace{12pt}compressed\_size\hspace{12pt}offset\hspace{12pt}filename>}

Of importance for this work are the URL, the timestamp, the status code, as well as the size. As the CDX files that we used for this analysis only include text files, such as HTML, we could ignore the mime type. Please note that the sizes provided in the CDX files corresponding to the records in the archive, compressed in \textit{GZip} format. Therefore, the analysis on sizes does not present the exact sizes of the websites, but trends over time.

In order to handle the large amount of data, we created an index based on the domains as keys. Each domain points to a list of its URLs, where every URL has attached a sub-list with all its captures in the archive in chronological order, including the data as shown above. This allows quick access to all URLs and captures of any available domain.

\subsubsection{Today's Popular Domains}
\label{sec:popular_domains}

There are three types of Web archives. While the first type attempts to preserve a certain part of the Web completely, for instance a national top-level domain, the second type is more focused, aiming for a certain topic or event. Those broad as well as topical crawls are typically done once or periodically without the attempt to capture all changes in between or to preserve the dynamics of the Web. In contrast to that, the third type of Web archives constitutes continuous crawls over a longer time period, which does not claim to preserve everything, but the most \text{important} parts according to a certain crawling strategy. This strategy might even change over time to adjust the crawler for a better coverage of a certain aspect. For instance, a typical strategy is to revisit frequently changing pages more often. Therefore, the temporal coverage of some websites in the archive may be very good, while others are missed completely. This selective crawling introduces a certain bias to the archive, which however is difficult to track retrospectively.

Our collection is of the third type, plus, it includes data donations, which were crawled by third-party organizations. For that reason, it does not cover the entire Web, but constitutes a sample biased by the different crawling strategies. Accordingly, a random sample of the collection would again be biased and it will require further research to analyze what the collection actually consists of to create a more representative sample of the entire German Web.

Therefore, instead of sampling we decided to focus on a well-defined subset, which in addition is inherently substantial for users as well as Web crawlers: the today's popular domains from their early stages in 1996 up to now (2013 to be exact). These websites are of interest for most readers and at the same time have the biggest impact on upcoming research on Web archiving, crawling, IR and related areas, since those disciplines typically focus on rather prominent websites. Also, as we will show later (s. Sec.~\ref{sec:dataset_analysis}), this subset nicely represents the actual growth of the Web in terms of registered domains. 

The selection of domains was taken from \textit{Alexa} by fetching the top websites of different categories, like \textit{Business}, \textit{Society}, \textit{Sports} and others. To match our dataset we only picked those categories listed under German~\footnote{\url{http://alexa.com/topsites/category/Top/World/Deutsch}}. In addition to the top categories, we also took two sub-categories for news and universities, which we considered especially relevant. As our dataset only consists of domains ending with the German top-level domain \textit{.de} and not all German websites listed on Alexa are under \textit{.de}, we filtered out those websites with another top-level domain. Out of the remaining, we picked the top 100 from every category (or less for smaller categories, like news) to form our dataset. The last time we retrieved the rankings from Alexa was on July, 10th 2014 at 09:26 GMT+1.

\subsubsection{Dataset extraction}

Based on the selected domains from Alexa, we filtered our CDX dataset by taking only those records with URLs belonging to one of the domains. Additionally, we cleaned the dataset by discarding the following URLs:
\begin{itemize}[noitemsep]
\item All URLs ending with one of the following extensions: \textit{.jpg, .png, .gif, .css, .js}, because these constitute embeds and not self-contained resources, like websites. Although the dataset only consists of URLs with mime type \textit{text}, it included image types either because the server returned a wrong type or the files were not available and pointed to an error page. 
\item All URLs that have never returned a successful HTTP status code (starting with 2). Those are most likely broken links, which the crawler followed, but which did not lead to a successful response.
\item All URLs that were not crawled anymore in 2013, i.e., the last year of the dataset, even if the last available capture was successful. Keeping them would result in an inconsistent state, because we cannot tell what happened to them after the last time they were crawled.
\item All URLs that have been crawled successfully only once, even if this was in 2013. As it exists only a single capture of those pages, they do not contribute to our evolution analysis at this point. Most likely, the Internet Archive crawler has just begun to crawl them.
\end{itemize}

Ultimately, we ended up with a dataset consisting of 17 categories with
today's popular domains from the German Web, as presented in
Table~\ref{tab:dataset_statistics}. The dataset covers in
total 1,444 domains with 5,846 sub-domains and more than 20 million
URLs (20,778,475 URLs to be exact).

\subsection{Statistics and Metrics}
\label{sec:statistics}

Our statistics were gathered in two steps. First, a precomputation step counted different properties of a domain. Afterwards, we aggregated these properties into meaningful metrics. The following subsections describe these two steps in detail and define the terminology used in the analysis results (Sec.~\ref{sec:age-analysis} and \ref{sec:url-analysis}). 

We use the terms of \textit{birth}, \textit{death} and \textit{life} to describe the lifetime of a URL or domain in our dataset. We consider a URL or domain to be alive from the time it first appeared in the Web archive until it was last seen online. 

\subsubsection{Precomputations}
\label{sec:precomputations}

For each domain, we precomputed three types of statistics: \textit{Evolution}, \textit{Domain Age} and \textit{URL Age} statistics. Each of them describes a collection of properties, such as \textit{size} and \textit{age}, computed in different units, i.e., calendar years, domain years, URL years. For all statistics, one unit $i$ spans a period $p_i$ of one year time from $t_{i}$ to $t_{i + 1}$ (excluded), which may or may not be a calendar year from 1 Jan to 31 Dec, depending on the type of statistics presented.

We decided not to collect more fine-grained statistics, such as monthly or weekly, because a higher resolution would not have had any advantages for our analysis and is not sufficiently supported by the dataset. While studies on change rates would require more steady crawls, this is not required for an evolution study such as the one we present as the overall trends are not affected. Also, we cannot guarantee such fine-grained captures with our dataset (s. Sec.~\ref{sec:dataset_analysis}).

The following definitions describe the statistics:
\begin{itemize}
\item \textbf{\textit{Evolution} statistics}:\\
Values are measured per calendar year.\\
$t_i$ denotes the beginning of the calendar year $i$.
\item \textbf{\textit{Domain Age} statistics}:\\
Values are measured for full years starting from the first date a
domain occurs in the dataset (e.g., for a domain that appears first in
$t_{0} =$ 04.05.2000 10:30:45, age $i =$ 0 spans from to
04.05.2001 10:30:44).\\ 
$t_i$ denotes the beginning of the domain age $i$.
\item \textbf{\textit{URL Age} statistics}:\\
Values are measured for full years of the analyzed URLs. As before the statistics are gathered per domain, however, here by combining values of different URLs at the same age.\\ 
$t_i$ denotes the beginning of the URL age $i$.
\end{itemize}

Age statistics (\textit{Domain Age} and \textit{URL Age}) do not
necessarily reflect the actual age of domains/URLs, but their age as evident from the dataset. These ages probably do not diverge much, but some time might have passed after the creation of a new domain until it is included in the Web archive. 

\textit{Evolution} and \textit{Domain Age} statistics are similar in
the sense that both describe the evolution of a domain over time. The
\textit{URL Age} statistics on the other hand are relative to the time
of a domain's URLs, which reflects different periods of a
domain but aggregates URLs at the same age. This enables different kinds of statistics as shown below.

\subsubsection{Aggregation}

The precomputed statistics were accumulated among the domains in each category as well as among all categories. For the sake of clarity, we present only selected categories in our plots, which best represent the overall observations as well as some outliers. Each metric that we analyze below is defined per period $p_i$ on the set of domains that appeared in this period $D_i$. For instance, a domain which was born in the year 2000 is not included in $D_i$ for any $i < 2000$ in the \textit{Evolution} statistics. The same applies to \textit{Domain Age} and \textit{URL Age} statistics with $i$ referring to relative years instead of calendar years.

The aggregations with corresponding formulas that we present and discuss in the following of this paper are presented along with the plots. The definitions of the used properties are listed in Table~\ref{tab:property_reference}. In addition to the given definition of alive URLs, we define the number of URLs alive at a single time point, which is a special case for a period with length 0: while $\text{alive}_d(p_i)$ is defined for a period $p_i = [t_i, t_{i+1})$ and denotes the URLs that were alive the entire interval, $\text{alive}_d(i)$ refers to the very end of this period. It includes the URLs that were alive during the entire period $p_i$ plus the ones that were born in period $p_i$: 
\begin{equation*}
\text{alive}_d(i) = \text{alive}_d(p_i) + \text{born}_d(p_i)
\end{equation*}

\section{The Age of the Web}
\label{sec:age-analysis}

The Web started around 25 years ago and has been maturing ever since. However, is its actual age really increasing or is its content constantly being refreshed, by pages being added and removed? To answer this question we analyzed the age of the Web in terms of how long URLs have been existent. It turns out, while the majority of popular webpages are young, older pages are aging further. We show the distribution of ages among URLs as well as the evolution of the long-living parts of the Web.

\begin{figure}[t]
\centering
\includegraphics[width=0.4\textwidth]{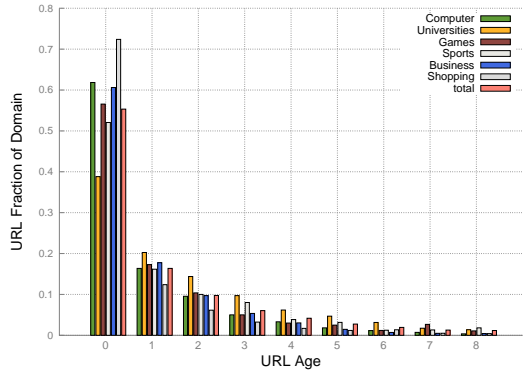}
\caption{URL Age Distribution}
\label{fig:age_distribution_per_domain}
\vspace{-0.2cm}
\end{figure}

\subsection{Distribution}
\label{sec:age_distribution}

It has been shown by other researchers that most URLs on the Web are rather short living~\citep{ntoulas_whats_2004,fetterly_large-scale_2003}, i.e., less than a year. However, nothing could be deduced about the URLs which survived after a year. Also, there was no evidence whether the fraction of these short-term URLs increased or decreased over time. To answer these questions, we first investigate the age distribution to determine what fraction of URLs is short or longer living and how this differs among the different categories. In this analysis, we only consider URLs that died during the timespan of our dataset, determined by an unsuccessful status code without another successful status code thereafter. The end time of such a URL is set to the time of the first returned unsuccessful status code. The begin time of the URL is the time it was crawled first. 

Figure~\ref{fig:age_distribution_per_domain} shows the fraction of URLs per domain that died at age $i$, averaged over all domains $D_i$ that reached this age. It is defined on the \textit{URL Age} statistics (cp. Sec.~\ref{sec:statistics}), with $p_i$ referring to the period of a URL's age $i$:
\begin{equation*}
\label{eq:age_distribution}
\frac{1}{|D_i|}\sum_{d \in D_i} \frac{\text{died}_d(p_i)}{\text{count}_d(p_i)}
\end{equation*}

The age distribution shows that, indeed, the largest fraction of the URLs of a domain, about 55\%, live less than a year. A considerable fraction of URLs die at the age of two to five. These are what we denote as short-living pages. Every page that lives longer than five years is considered long-living and subject to contribute to the aging of the Web. These constitute the long tail in this distribution.  We do not show the entire tail in this figure but we considered URLs up to ages of thirteen. It is interesting to observe that the university websites have a significantly higher number of URLs dying after the first year, while less than 40\% of webpages die at the age of 0. For each of the subsequent ages they consistently outnumber other categories indicating that university webpages tend to be rather long-living. In contrast, we have shopping websites, which have the highest number of pages, 73\% of all its URLs, that die within their first year.

\begin{figure*}[t]
    \centering  
    
    \subfigure[Total] {\label{fig:age_0to5_per_year}\includegraphics[width=0.4\textwidth]{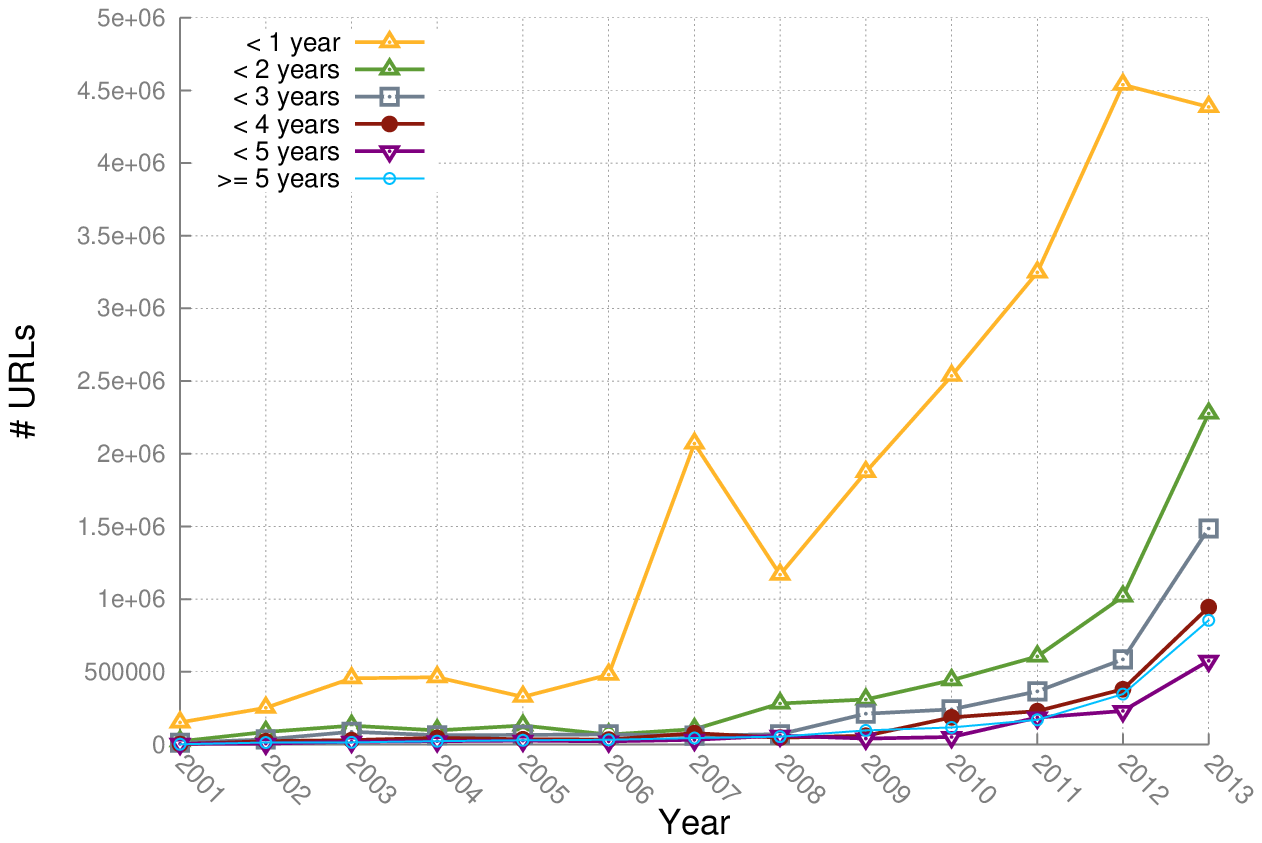}}
    \hspace{1cm}
    \subfigure[Normalized] {\label{fig:age_per_year_normalized}\includegraphics[width=0.4\textwidth]{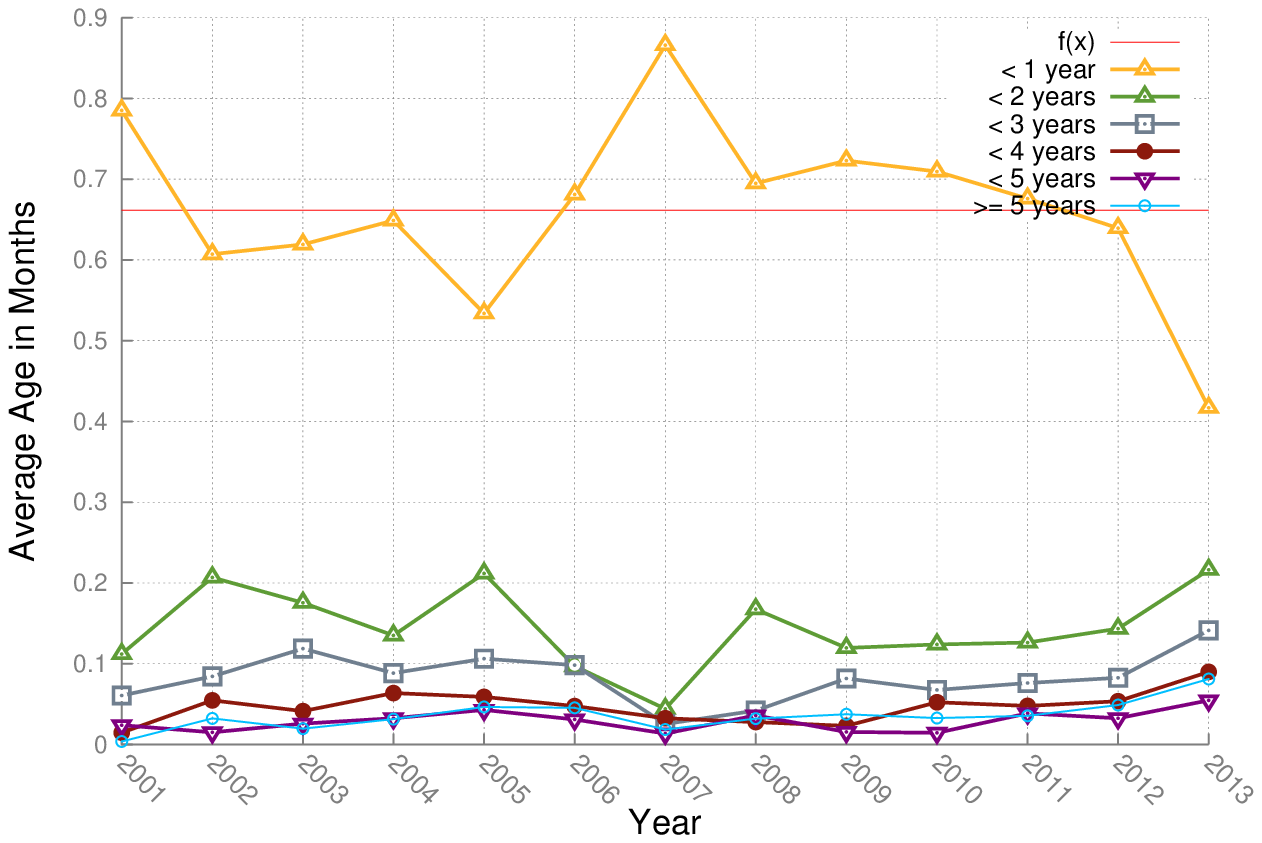}}

    \caption{Evolving URL Age Distribution}
    \label{fig:lag-dist-low-med-high}
    \vspace{-0.2cm}
\end{figure*}

Now we turn to the second question of how the overall age distribution evolves over time, presented in Figures~\ref{fig:age_0to5_per_year} and~\ref{fig:age_per_year_normalized}. For this, we resort to a different style of analysis by considering the number of URLs at a certain age in the given year, instead of how long they lived in the end. We divided the ages into six age buckets of URLs that lived for less than -- a year, 2 years, 3 years, 4 years, 5 years and 6 years or longer, which includes the URLs at age five together with the long-living ones. We observe in Figure~\ref{fig:age_0to5_per_year}, that over the years the number of URLs for each bucket increases superlinearly. Interestingly, this trend correlates with the domain volume which is presented in the next section. 

Further, we investigate the normalized distribution for all years in Figure~\ref{fig:age_per_year_normalized}. The normalized value of an age bucket $\alpha$ at a given year $p_i$ is defined as follows (on \textit{Evolution} statistics):
\begin{equation*}
\label{eq:age_fraction}
\frac{\sum_{d \in D_i}|\{a \in \text{ages}_d(p_i) | \alpha \cdot 12 \leq a < (\alpha + 1) \cdot 12 \}|}{\sum_{d \in D_i}\text{alive}_d(i)}
\end{equation*}

Although the number of URLs overall grows over the years, as suggested by
Figure~\ref{fig:age_0to5_per_year}, the fraction of the URLs at
different ages remains more or less stable. As emphasized by the
computed fitted line in Figure~\ref{fig:age_per_year_normalized},
almost 70\% of all webpages are younger than a year at any time during
the Web's lifetime. The fact that the sizes of all age buckets are equally stable over time suggests that, although the Web is growing, it consists of equal proportions of different aged webpages at any time.

As a result of the retrospective nature of this study, abnormal artifacts that appear in some of the plots are difficult to track. Similar to the peak in year 2007 in Figure~\ref{fig:age_0to5_per_year} there are artifacts in the following figures as well. These kinds of abnormalities are most likely due to the different data sources that donated crawls of very diverse volume and size to the Internet Archive. However, as all of them are local phenomena, they do not affect our analysis as the global trend can be clearly recognized in all figures. More details on the dataset are discussed in Section~\ref{sec:dataset_analysis}.

\subsection{Aging}

\begin{figure}[t]
\centering
\includegraphics[width=0.4\textwidth]{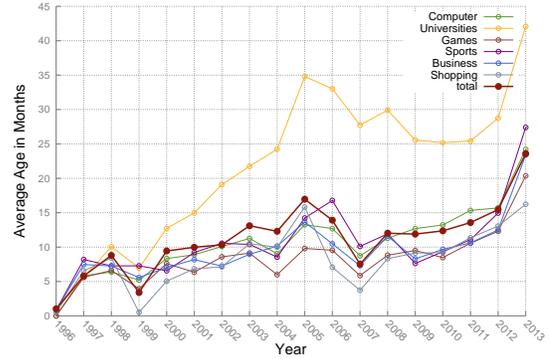}
\caption{URL Age Evolution}
\label{fig:average_age_month_evolution}
\end{figure}

Knowing that the majority of pages on the Web are rather fresh, we now
analyze the evolution of the Web's average age. Rather ironically,
like humans can grow old but stay younger by eating healthy and
doing sports, a similar trend applies to the Web as most of its
constituent webpages are frequently replaced. To investigate this, we
computed the \emph{average age} of the Web in months at any given year as defined below (on \textit{Evolution} statistics) and plotted in
Figure~\ref{fig:average_age_month_evolution}:
\begin{equation*}
\label{eq:average_age}
\frac{\sum_{d \in D_i}\sum_{a \in \text{ages}_d(p_i)} a}{\sum_{d \in D_i} |\text{ages}_d(p_i)|}
\end{equation*}

The figure shows that the Web is actually growing older after
all. While the average age of the Web was about 10 months during the
year 2000, it grew almost 50\% by the year 2012. This can
possibly be attributed to the stability of age distributions as shown
before
(s. Figure~\ref{fig:average_age_month_evolution}). Specifically, the
fraction of long-living webpages, which are constantly aging,
contributes to a higher age every year.

This aging is almost linear, following the curve $f(x) = a \cdot x +
b$, where $x$ is the number of years calculated from 1996. The
estimated values for the parameters of this curve are $a = 0.74, b =
4.89$ with an asymptotic error of $8.41\%$ (the corresponding plot is
attached in Figure~\ref{fig:fit_average_age_month_evolution}). This aging would
lead to an average URL age of 23 month in the year 2020, which is
double the age of 2005, while the age today or at the end of our dataset (2013) to be exact is 1.5 years. According to this finding, the Web will on average turn three in 2038. However, as our dataset goes back only until 1996, there might be even older pages on the Web. For this reason, our result can be considered as a lower bound.

We further verify our claim that this aging is caused by the
long-living pages by analyzing the age of webpages older than five
years using the following expression (defined on \textit{Evolution}
statistics):
\begin{equation*}
\label{eq:long_living_average}
\frac{\sum_{d \in D_i}\sum_{a \in \{a \in \text{ages}_d(p_i) | a > 5 \cdot 12 \}} a}{\sum_{d \in D_i} |\{a \in \text{ages}_d(p_i) | a > 5 \cdot 12 \}|}
\end{equation*}

\begin{figure*}[t]
    \centering  
    
    \subfigure[Evolution] {\label{fig:long_tail_average_per_year}\includegraphics[width=0.4\textwidth]{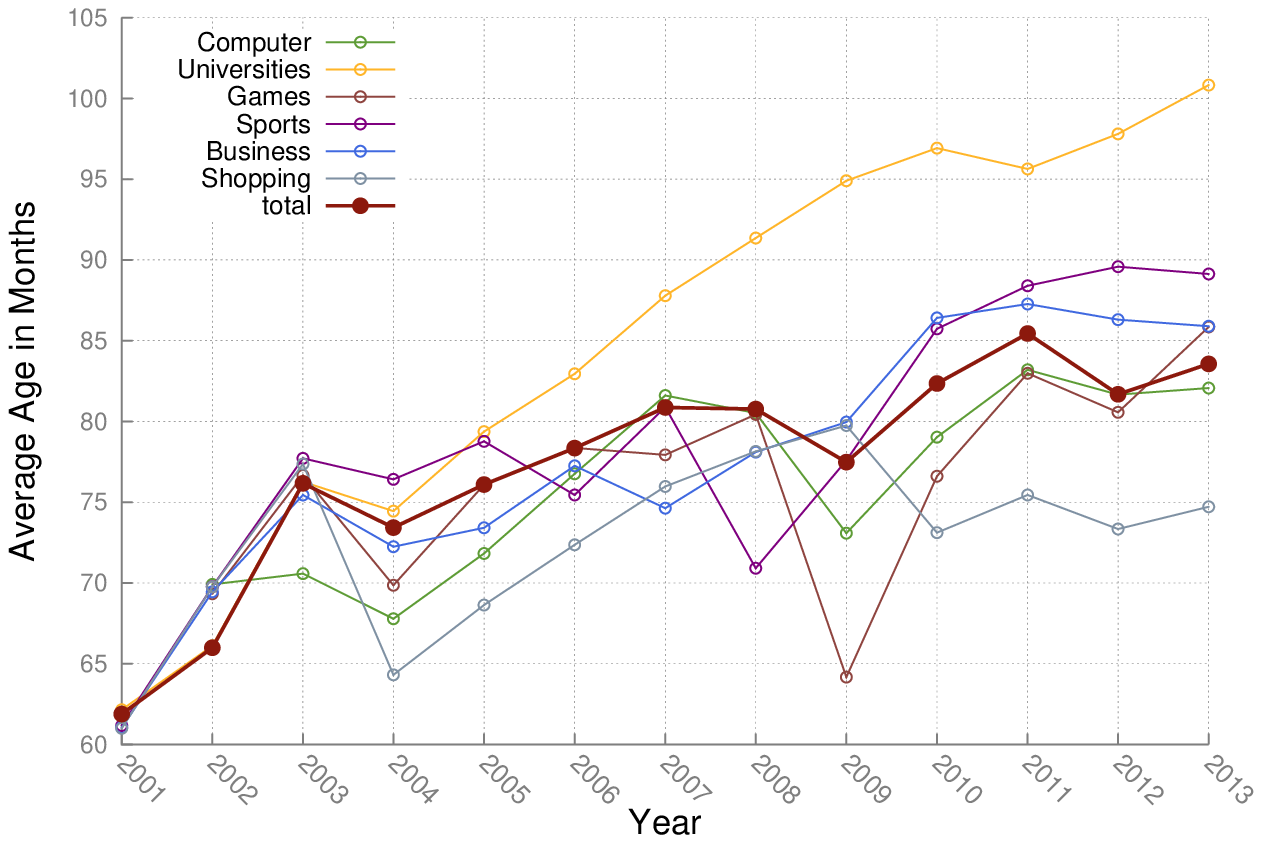}}
    \hspace{1cm}
    \subfigure[Domain Life] {\label{fig:long_tail_average_per_domain_age}\includegraphics[width=0.4\textwidth]{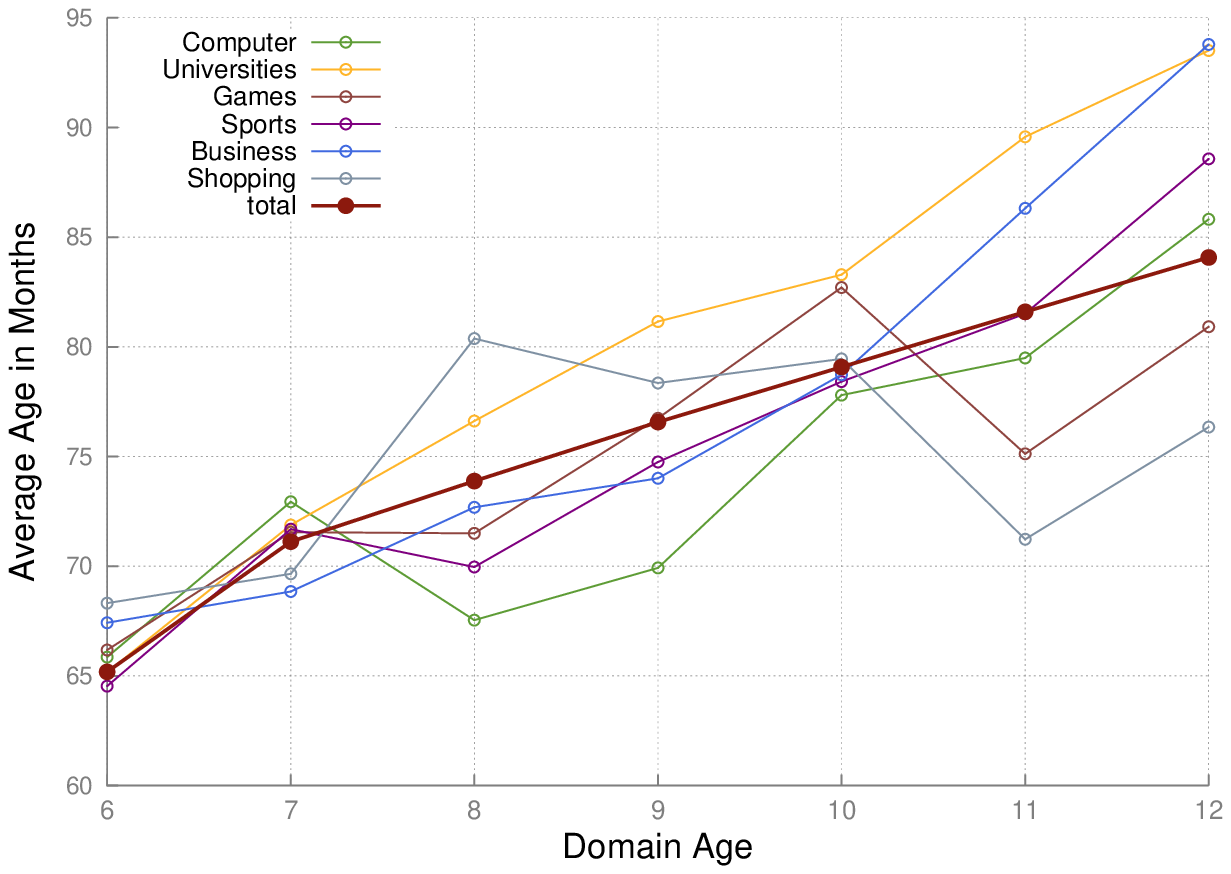}}

    \caption{Age of Long-Living URLs (older than five years)}
    \vspace{-0.2cm}
\end{figure*}

The corresponding plot in Figure~\ref{fig:long_tail_average_per_year}
visualizes the quite significant growth in age of the long-living
URLs. Even though this old part is just a small fraction of the entire
Web, its increasing age leads to the slow increase of the Web's actual
age that we have shown above. This figure only starts in 2001 as there
exist no long-living URLs in our dataset before.

The same observation can be made by analyzing the average age of
long-living URLs at a given age of the corresponding domains in
Figure~\ref{fig:long_tail_average_per_domain_age}. This is defined by
the same formula as used before, but on \textit{Domain Age}
statistics with $p_i$ referring to the of age $i$ of a domain (cp. Sec.~\ref{sec:statistics}). The plot reflects the actual
aging of the popular domains in our dataset in contrast to their real age, as shown on the x-axis: when a domain turns 10 years, their URLs are on average only 80 months old, which is about 6.5 years.

Corresponding to what we observed in
Section~\ref{sec:age_distribution} all plots in this section
acknowledge the characteristics in terms of age for
different categories. While websites of universities appear to be the
oldest, others such as sports, business and computer websites tend to
be much fresher, not to say more up to date.

\begin{figure}[t!]
\centering
\includegraphics[width=0.4\textwidth]{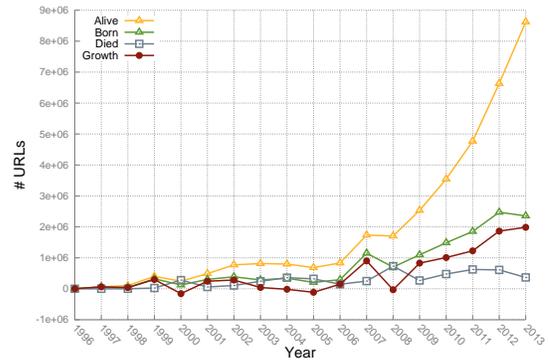}
\caption{Evolution of the Web's URL Volume}
\label{fig:number_of_urls_per_year}
\vspace{-0.2cm}
\end{figure}

\pagebreak
\section{The Growth of the Web}
\label{sec:url-analysis}

We now turn our attention to measuring the \emph{size of the popular Web} and how it has evolved over time. The size of the Web can be interpreted as the number of webpages or as the actual size of its content. We refer to the number of websites and pages as the volume of the Web or a domain, while size refers to the actual file size (including markup as well as the content of a page). In this section we study both interpretations and their evolution over time.

By design, we expect growth as we focus on today's popular domains, which have grown popular over time and therefore, have naturally grown in volume and probably size, too. The question now is how this growth, which made the websites as popular as they are today, can be characterized.

\subsection{Volume}

\begin{figure*}[t]
    \centering  
    
    \subfigure[Domain Volume Evolution] {\label{fig:number_of_urls_per_domain_per_year}\includegraphics[width=0.4\textwidth]{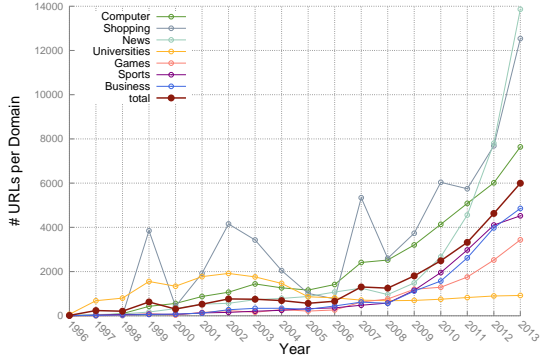}}
    \hspace{1cm}
    \subfigure[Birth/Death/Growth Rates Evolution] {\label{fig:birth_death_rate_per_domain_year}\includegraphics[width=0.4\textwidth]{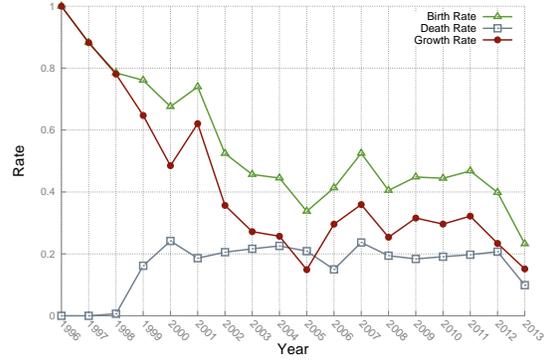}}
    
    \subfigure[Domain Volume over Domain Life] {\label{fig:number_of_urls_per_domain_age}\includegraphics[width=0.4\textwidth]{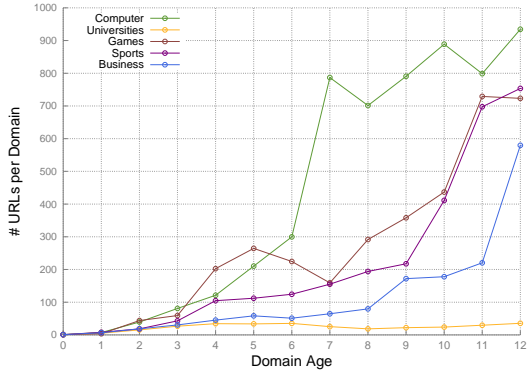}}
    \hspace{1cm}
    \subfigure[Birth/Death/Growth Rates over Domain Life] {\label{fig:birth_death_rate_per_domain_age}\includegraphics[width=0.4\textwidth]{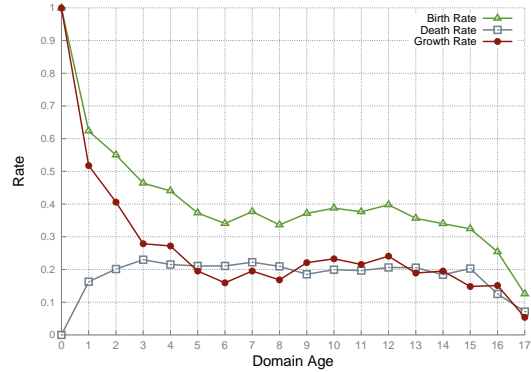}}

    \caption{Domain Volume}
    \vspace{-0.2cm}
\end{figure*}

Considering that the number of domains in our dataset grows every year as we will see in Section~\ref{sec:dataset_analysis}, it is not surprising that the number of URLs grows as well. However, if this was the only reason, the growth would be similar to the growth of our dataset, which is not the case. We analyzed this by computing four properties: (a) the number of newborn URLs in a year, (b) the number of URLs that died in a year, (c) the number of URLs that are alive at the end of a year, as well as (d) the growth rate. The growth rate is the difference between the number of born and died URLs. While all other numbers are computed over the period of a year $p_i$, the number of URLs alive is considered at the end of the year $i$, defined as follows (on \textit{Evolution} statistics):
\begin{equation*}
\sum_{d \in D_i} \text{alive}_d(i)
\end{equation*}

The results are presented in Figure~\ref{fig:number_of_urls_per_year}, which shows that the Web is growing a little faster every year. Especially noticeable is the strong growth starting from 2006, which however might be due to the characteristics of the dataset after all. The reason for this growth of the Web is that there are more new URLs born every year, while the number of dying URLs remains almost constant. In order to affirm that this finding is independent from the growing number of domains in our dataset, we investigated the average number of URLs per domain over the years as well. The formula below (defined on \textit{Evolution} statistics) describes this progression, which is shown by the plots in Figure~\ref{fig:number_of_urls_per_domain_per_year} per category:
\begin{equation*}
\label{eq:alive_per_domain}
\frac{1}{|D_i|} \sum_{d \in D_i} \text{alive}_d(i)
\end{equation*}

Figure~\ref{fig:birth_death_rate_per_domain_year} shows the corresponding average growth rate per domain, as defined below (on \textit{Evolution} statistics), together with birth and death rates. The growth rate describes the difference of born and died URLs of one domain in a given year as fraction of the ones that were alive at the beginning of the year:
\begin{equation*}
\label{eq:growth_rate}
		\frac{1}{|D_i|} \sum_{d \in D_i}\frac{\text{born}_d(p_i) - \text{died}_d(p_i)}{\text{alive}_d(p_i) + \text{died}_d(p_i)}
\end{equation*}

Except for the beginning of this plot, which is most likely due to the transient state at the early years of our dataset, the growth rate is relatively stable at around 30\%. Based on this, we can deduce that the number of URLs that are born or die depends on the volume of the Web or their domain. However, among categories the growth varies strongly. While most of them follow the overall trend, university websites barely grow in volume at all, as presented earlier in Figure~\ref{fig:number_of_urls_per_domain_per_year}. Even in 2013 they still only consist of about 1,000 URLs on average, whereas computer websites comprise almost 8,000 and shopping as well as news websites more than 12,000 URLs.

The average domain volume follows an exponential curve $f(x) = a \cdot b^x + c$, where $x$ is the number of years calculated from 1996. The estimated values for the parameters of this curve are $a = 22.82, b = 1.38, c = 300.18$ with an asymptotic error of $2.07\%$ (the corresponding plot is attached in Figure~\ref{fig:fit_number_of_urls_per_domain_per_year}). Assuming the growth continues with the same rate, in the year 2020 the number of URLs of the popular domains would be almost 6.7 times the number of URLs today (2014) and by 2030 it would be 166 times that of today. Already within the next two years the domain volume would be doubled. Even though this prediction might be weakened due to our crawling assumptions for archives (s. Sec~\ref{sec:dataset_analysis}) or the resource limiting is not exponential with the same degree (which is indeed the case as confirmed by the Internet Archive), the exponential nature is still retained, although not as strong.

Another perspective to look at the growth of websites is from the age of a domain in contrast to absolute years. Instead of plotting total numbers, this time we analyzed the number of URLs at every age of a domain in relation to its initial volume (defined on \textit{Domain Age} statistics):
\begin{equation*}
\label{eq:relative_domain_volume}
	\frac{1}{|D_i|}\sum_{d \in D_i} \frac{\text{alive}_d(i)}{\text{alive}_d(0)}
\end{equation*}

Figure~\ref{fig:number_of_urls_per_domain_age} gives an impression of this relative volume over the lifetime of a domain for five selected categories. We decided to look only at the first 12 years, as our data is not representative enough for older domains. Most noticeable is a quick growth at some point for the websites in most categories. However, the time of this critical take off varies. While computer websites appear to have a strong growth already very early around year six, where they reach 800 times the volume that they started with at birth, and stagnate afterwards, most categories take longer. As observed before, university websites hardly grow in volume at all. Interestingly, the average growth during the lifetime of a domain, as presented in Figure~\ref{fig:birth_death_rate_per_domain_age}, looks very similar to the actual growth of the popular German Web over time.

\subsection{Size}

\begin{figure*}[!t]
    \centering  
    
    \subfigure[Alive Size] {\label{fig:url_size_per_year}\includegraphics[width=0.4\textwidth]{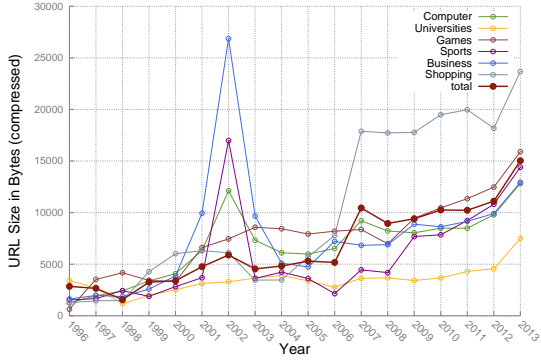}}
    \hspace{1cm}
    \subfigure[Birth Size] {\label{fig:new_born_size_per_year}\includegraphics[width=0.4\textwidth]{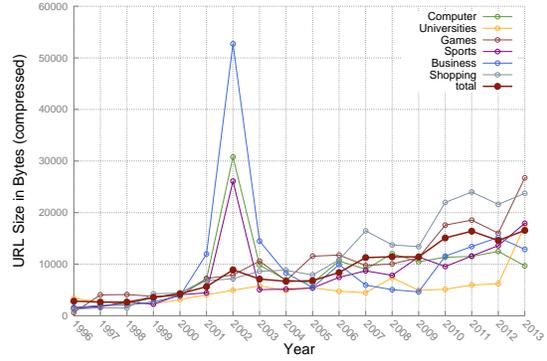}}

    \caption{URL Size Evolution}
    \vspace{-0.2cm}
\end{figure*}

Apart from the volume also the actual size in bytes has been growing. We found this to be the result of two evolutions: newborn URLs appear to be larger nowadays than they used to be earlier and, in addition, URLs grow in size during their lifetime. 

We first analyzed the average size of a URL evolving over time (defined on \textit{Evolution} statistics):
\begin{equation*}
\label{eq:url_size_evolution}
\frac{\sum_{d \in D_i}\text{size}_d(p_i)}{\sum_{d \in D_i}\text{alive}_d(i)}
\end{equation*}

Figure~\ref{fig:url_size_per_year} shows that the size of URLs indeed has increased over the years. This can either mean that websites today consist of more content than they used to in earlier days of the Web, or the markup has grown.

As it turns out, a major growth in size is contributed by newborn URLs, as defined below (on \textit{Evolution} statistics):
\begin{equation*}
\label{eq:size_at_birth}
\frac{\sum_{d \in D_i}\text{born\_size}_d(p_i)}{\sum_{d \in D_i}\text{born}_d(p_i) + \text{flashed}_d(p_i)}
\end{equation*}

\begin{figure*}[!ht]
    \centering  
    
    \subfigure[URL Age Evolution] {\label{fig:fit_average_age_month_evolution}\includegraphics[width=0.3\textwidth]{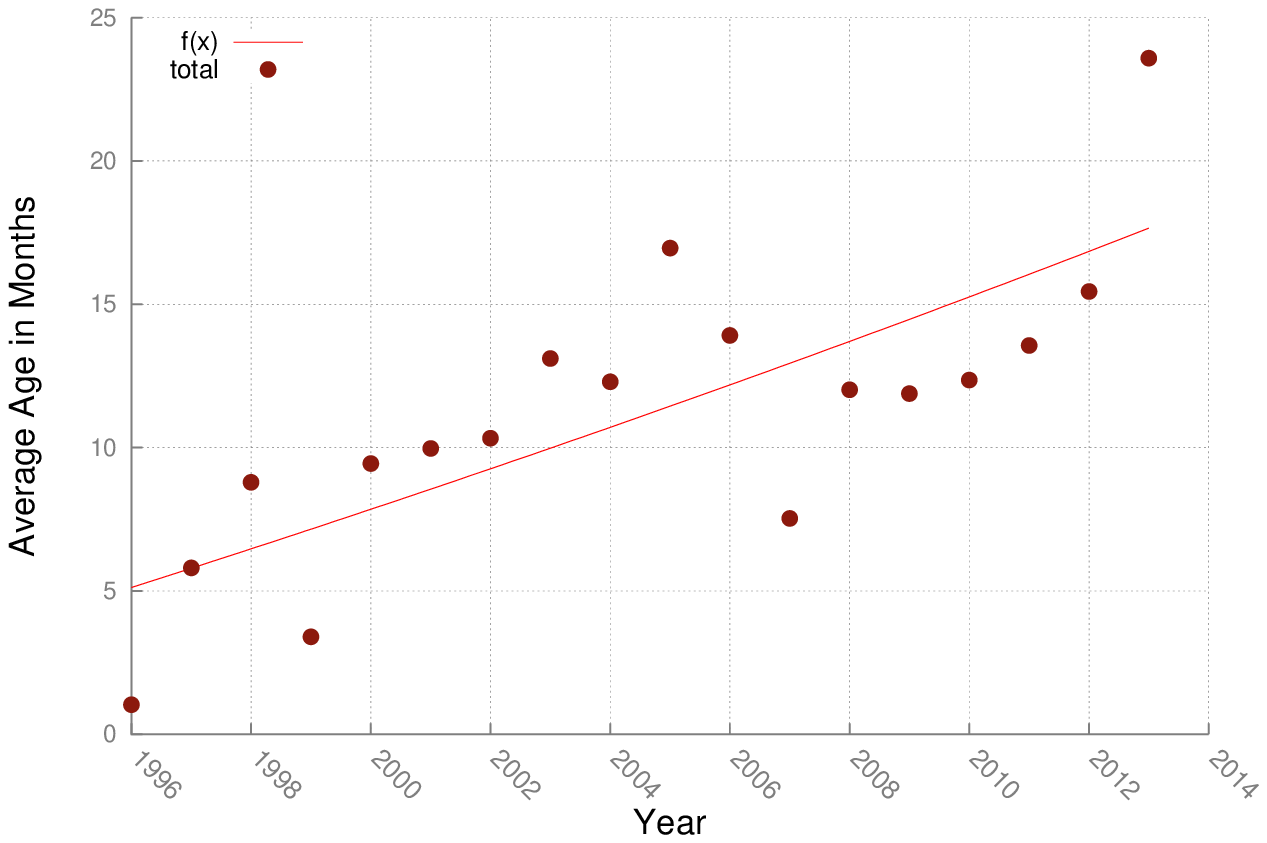}}
    \hspace{1mm}
    \subfigure[Domain Volume Evolution] {\label{fig:fit_number_of_urls_per_domain_per_year}\includegraphics[width=0.3\textwidth]{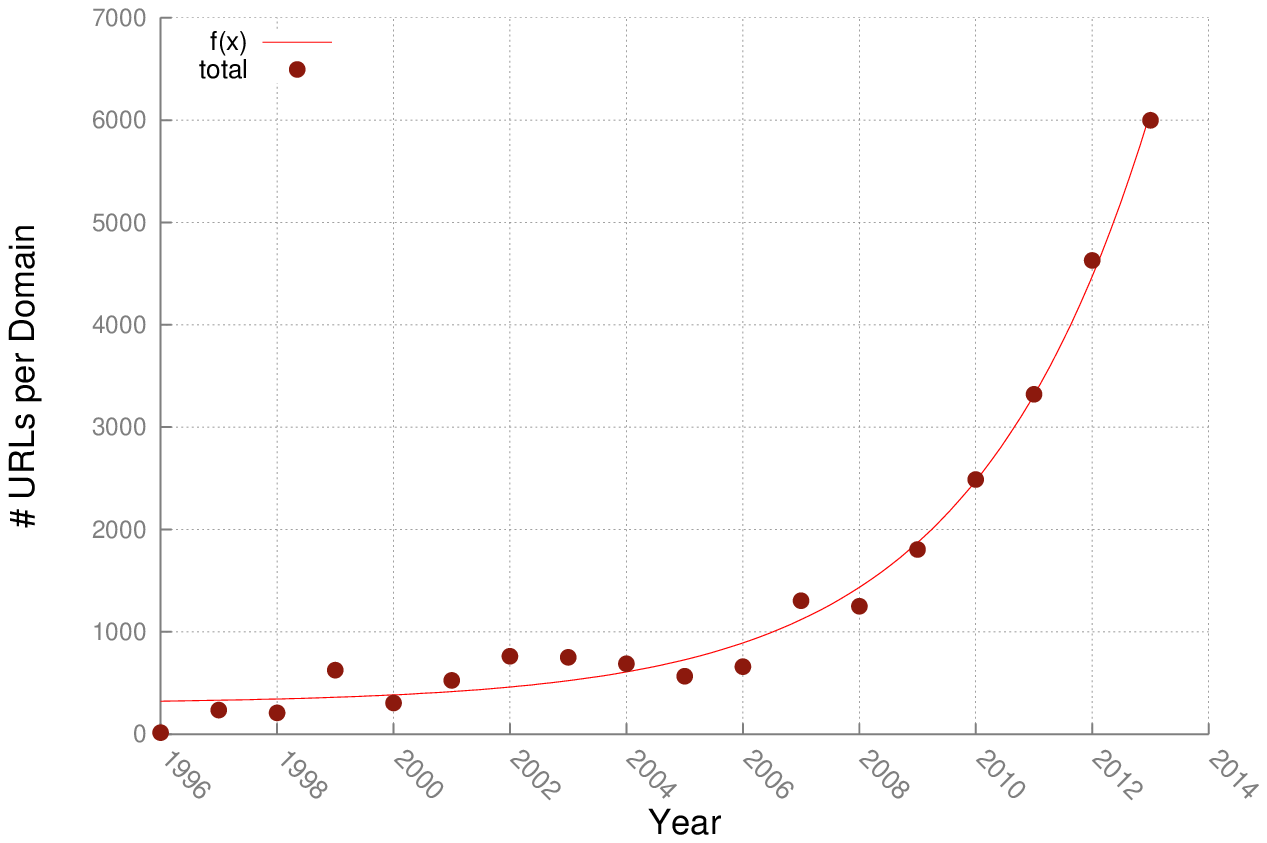}}
  \hspace{1mm}
    \subfigure[URL Birth Size Evolution ] {\label{fig:fit_new_born_size_per_year} \includegraphics[width=0.3\textwidth]{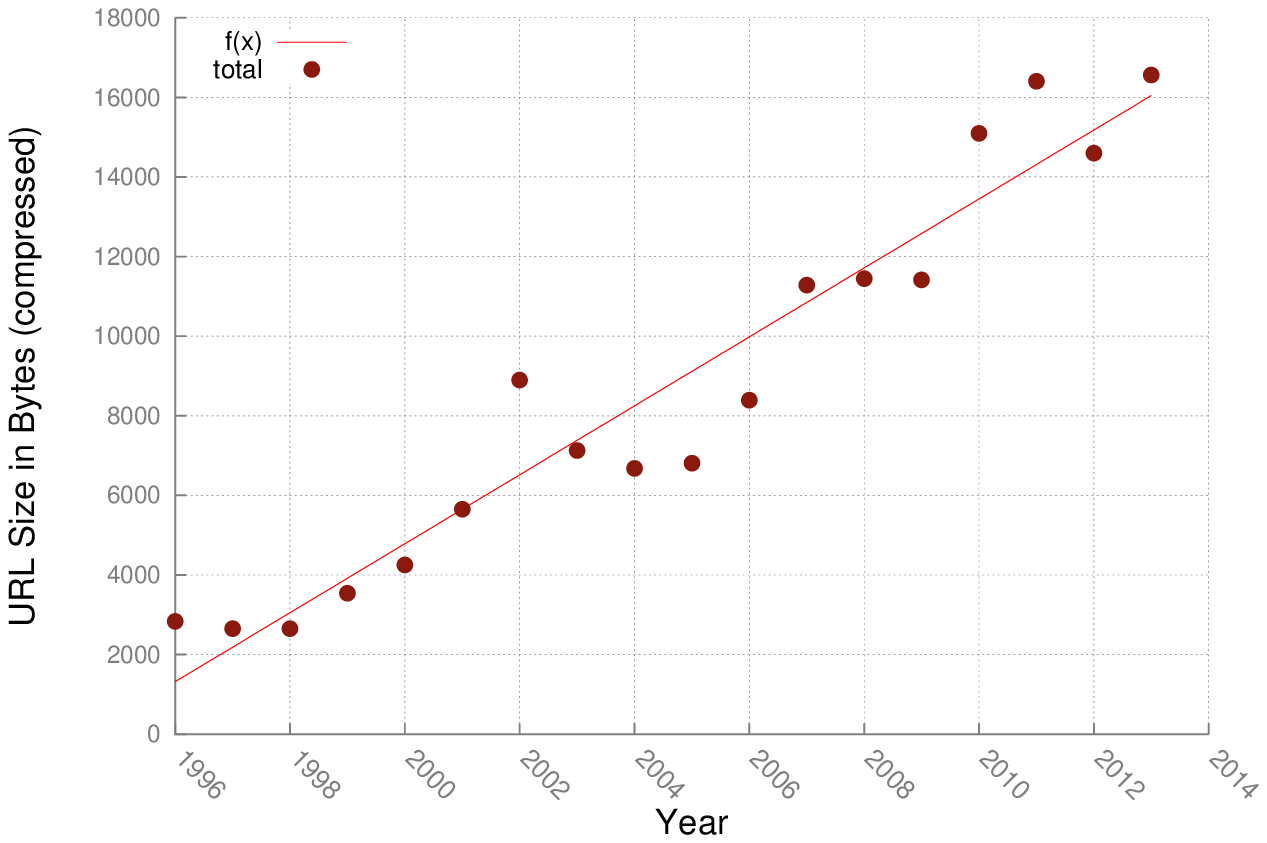}}
    \caption{Predictions of Evolution Analysis}
    \vspace{-0.2cm}
\end{figure*}

This evolution, presented by Figure~\ref{fig:new_born_size_per_year}, is similar to the overall growth in size. Its trend follows a linear curve $f(x) = a \cdot x + b$, where $x$ is the number of years calculated from 1996. The estimated values for the parameters of this curve are $a = 866, b = 1320$ with an asymptotic error of $6.9\%$ (the corresponding plot is attached in Figure~\ref{fig:fit_new_born_size_per_year}). Based on this, in the year 2038 a new URL will be born on average with double the size as today (2016). As these are compressed sizes (s. Sec.~\ref{sec:cdx}), we cannot state actual numbers though.

Another factor that contributes to the growth of URL sizes is the growth of existing URLs during their lifetime. For this analysis we only took those URLs into account that died at some point within the period of our dataset and computed the average size at birth and at death of all URLs that reached a certain age, as defined by the formulas below (on \textit{URL Age} statistics):
\begin{equation*}
\label{eq:cumulated_size_at_birth}
\frac{\sum_{d \in D_i, j \geq i}\text{died\_birth\_size}_d(p_j)}{\sum_{d \in D_i, j \geq i}\text{died}_d(p_j)}
\end{equation*}
\begin{equation*}
\label{eq:cumulated_size_at_death}
\frac{\sum_{d \in D_i, j \geq i}\text{died\_size}_d(p_j)}{\sum_{d \in D_i, j \geq i}\text{died}_d(p_j)}
\end{equation*}

\begin{figure}[t]
\centering
\includegraphics[width=0.4\textwidth]{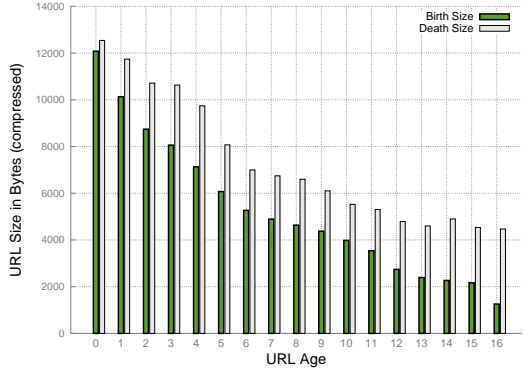}
\caption{Average URL Birth/Death Size}
\label{fig:birth_death_sizes_for_different_ages}
\vspace{-0.2cm}
\end{figure}

Figure~\ref{fig:birth_death_sizes_for_different_ages} shows these numbers in a cumulative manner, averaged over all URLs at a given age. Accordingly, URLs that die earlier tend to be larger than longer living ones. Hence, it appears that less content promises a longer lifetime. Furthermore, the plot shows that URLs grow in size over time, regardless of their age. This growth is almost constant, which indicates that longer living URLs either grow more slowly or that most of the growth takes place in the early years of a URL, as already found by Koehler et. al~\cite{koehler_web_2002}. In contrast to that observation, short-living URLs with a lifetime of less than a year seem to grow least of all in size.

\section{Archive Dataset Discussion}
\label{sec:dataset_analysis}

Our analysis of Web evolution is performed on a dataset comprising German websites under the \textit{.de} top-level domain, which was provided by the Internet Archive. The Internet Archive is the largest and most complete Web archive today. It covers a period of 18 years and constitutes a great source for analysis like ours. Just like in every other archive, not everything can be preserved. What is saved from the Web is influenced by crawl policies and constraints that impact both completeness as well as the change coverage.

We conducted this analysis under the major assumption that, if a domain is crawled, it is crawled completely with respect to the applied crawling policies and limitations, such as certain filters and maximum number of hops from a seed page. Hence, even though this does not cover all URLs of a domain, as long as the crawling strategy does not change over time, our observed trends are still valid. For Internet Archive crawls performed after 2010 this is actually the case. Thus, at least our results after that time are not affected by changing crawl policies at all. However, due to our focus on popular domains, we expect the assumption to be widely true also before 2010. The Internet Archive received lots of their crawls as donations from different partners. As crawlers, especially from search engines, typically aim for the most prominent part of the Web, we consider our subset consisting of popular domains to be covered with higher priority and hence very comprehensively compared to the rest of the archive.

\begin{figure}[t]
\centering
\includegraphics[width=0.4\textwidth]{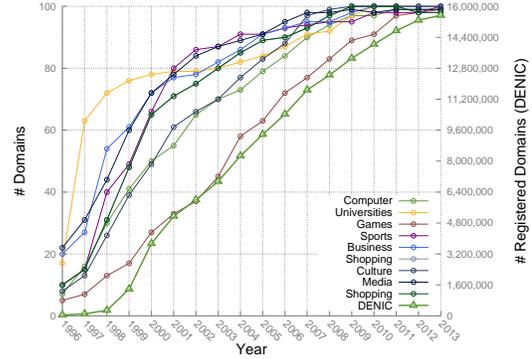}
\caption{Domain Emergence vs. Registered Domains on DENIC (right y-axis)}
\label{fig:number_of_domains_per_year}
\vspace{-0.4cm}
\end{figure}

Moreover, we investigated how well the analyzed popular domains in the Web archive represent the actual Web by comparing to the trend of registered domains on DENIC\footnote{\url{http://www.denic.de/en/background/statistics.html}} (the \textit{.de} domain registrar), as shown in Figure~\ref{fig:number_of_domains_per_year}. The plot gives an overview of presence of domains from the different categories in our dataset at every year under consideration. A domain that is not present can mean two things: 1. it was not online at that time, or 2. it was not considered in the Internet Archive crawls. Although we are not able to distinguish this, the experiment shows a similar trend to the actually existing domains, suggesting that our dataset is fairly representative.

Interesting are also the differences among different categories: whereas about 75\% of today's university websites already existed in 1999 and grew quickly, not even 20\% of today's popular game websites were present back then. Most likely, many universities even had a website before 1996, but only got picked up by the crawlers later. By contrast the game websites that are most popular today have been created more recently and grown slowly since. The fact that perhaps not all domains were covered in the very beginning does not affect our analysis, as we investigated volume and size on a per-domain and per-URL basis, respectively.

\section{Conclusion and Outlook}
\label{sec:conclusion}

In this paper, we have presented an extensive longitudinal study on 18 years of the popular German Web, based on crawls of the Internet Archive. We carried out an in depth analysis on how the popular domains of today were created and how their age, volume and sizes have grown over the last decade. First, we find that most of the popular educational domains like universities have already existed for more than a decade. On the other hand, domains relating to shopping and games have emerged steadily over the period of the last decade. Second, we see that the Web is getting older, not in all its parts, but with many domains having a constant fraction of webpages that are more than five years old and aging further. Finally, we see that popular websites have been growing exponentially after their inception, doubling in volume every two years.

The study has provided us with interesting insights and ramifications on the evolution of the prominent part of the German Web. What we have learned about its growth and size can impact resource allocation strategies for Web archives as well as exhaustive and focused crawling strategies. Especially the identified differences among the studied categories can be of importance when dealing with topical or organizational Web archives from the respective areas. The introduced properties and definitions provide a solid foundation for comparing our findings on growth and aging against different Web archive collections. A possible research question would be: How does the Web of other countries compare to this analysis of the German Web? Furthermore, we lay the foundation for follow-up questions in future research, such as: How do webpages evolve content-wise compared to size and age, and why is the average size of the newborn webpages today larger than the ones in the yesteryear? Is it because of an actual increase in content or is it because of the markup due to constantly increasing web authoring technologies? 



\bibliographystyle{unsrtnat}
\bibliography{references}

\end{document}